%% file: main.tex
\documentclass[10pt,twocolumn,letterpaper]{article}

\usepackage{cvpr}              
\usepackage{float}
\usepackage[symbol]{footmisc}
\usepackage{graphicx}
\usepackage{pifont}
\newcommand{\cmark}{\ding{51}}%
\newcommand{\xmark}{\ding{55}}

\usepackage[accsupp]{axessibility}
\input{preamble}

\definecolor{cvprblue}{rgb}{0.21,0.49,0.74}
\usepackage[pagebackref,breaklinks,colorlinks,citecolor=cvprblue]{hyperref}

\def\paperID{*****} 
\def\confName{CVPR}
\def\confYear{2024}

\title{%
Gene-Level Representation Learning via Interventional Style Transfer in Optical Pooled Screening}

\author{
Mahtab Bigverdi\,\textsuperscript{1}
$\footnotemark[1]${}
\quad Burkhard H\"ockendorf\,\textsuperscript{2}
$\footnotemark[1]${}
\quad Heming Yao\,\textsuperscript{2} \\
\quad Phil Hanslovsky\,\textsuperscript{2}
\quad Romain Lopez\,\textsuperscript{3,4}
\quad David Richmond\,\textsuperscript{2}\medskip\\
\textsuperscript{1}\,University of Washington 
\quad\textsuperscript{2}\,BRAID, Genentech 
\quad\textsuperscript{3}\,gRED, Genentech 
\quad\textsuperscript{4}\,Stanford University\\
\textsuperscript{1}\,{\tt\small mahtab@cs.washington.edu}
\quad \textsuperscript{2}\,{\tt\small \{hoeckendorf.burkhard, richmond.david\}@gene.com}
}

\AtBeginShipoutNext{\AtBeginShipoutUpperLeft{%
  \put(\dimexpr\paperwidth-2.3cm\relax,-1.5cm){\makebox[0pt][r]{\textcolor{gray}{This paper has been accepted for publication at the
IEEE Conference on Computer Vision and Pattern Recognition (CVPR)}}}%
}}

\AtBeginShipoutNext{\AtBeginShipoutUpperLeft{%
  \put(\dimexpr\paperwidth-5cm\relax,-2cm){\makebox[0pt][r]{\textcolor{gray}{
Workshop on Computer Vision for Microscopy Images (CVMI), Seattle, 2024. ©IEEE}}}%
}}

\begin{document}
\maketitle
\input{sec/0_abstract}    
\input{sec/1_intro}
\input{sec/related_work}

\input{sec/2_method}
\input{sec/3_result}
\input{sec/ablations}
\input{sec/4_conclusion}
{
    \small
    \bibliographystyle{ieeenat_fullname}
    \bibliography{main}
}


\end{document}

%% file: preamble.tex
\usepackage[dvipsnames]{xcolor}

\newcommand{\myparagraphn}[1]{%
\noindent\textbf{#1}\enspace}%
\newcommand{\myparagraph}[1]{%
\smallskip\myparagraphn{#1}}%

\usepackage{moresize}

%% file: sec/0_abstract.tex
\begin{abstract}
Optical pooled screening (OPS) combines automated microscopy and genetic perturbations to systematically study gene function in a scalable and cost-effective way. Leveraging the resulting data requires extracting biologically informative representations of cellular perturbation phenotypes from images. We employ a style-transfer approach to learn gene-level feature representations from images of genetically perturbed cells obtained via OPS. Our method outperforms widely used engineered features in clustering gene representations according to gene function, demonstrating its utility for uncovering latent biological relationships. This approach offers a promising alternative to investigate the role of genes in health and disease.
%

\end{abstract}

%% file: sec/1_intro.tex
\section{Introduction}
\label{sec:intro}

\footnotetext[1]{Equal contribution}

%
%
\begin{figure*}[t]
    \centering
    \includegraphics[width=.85\textwidth]{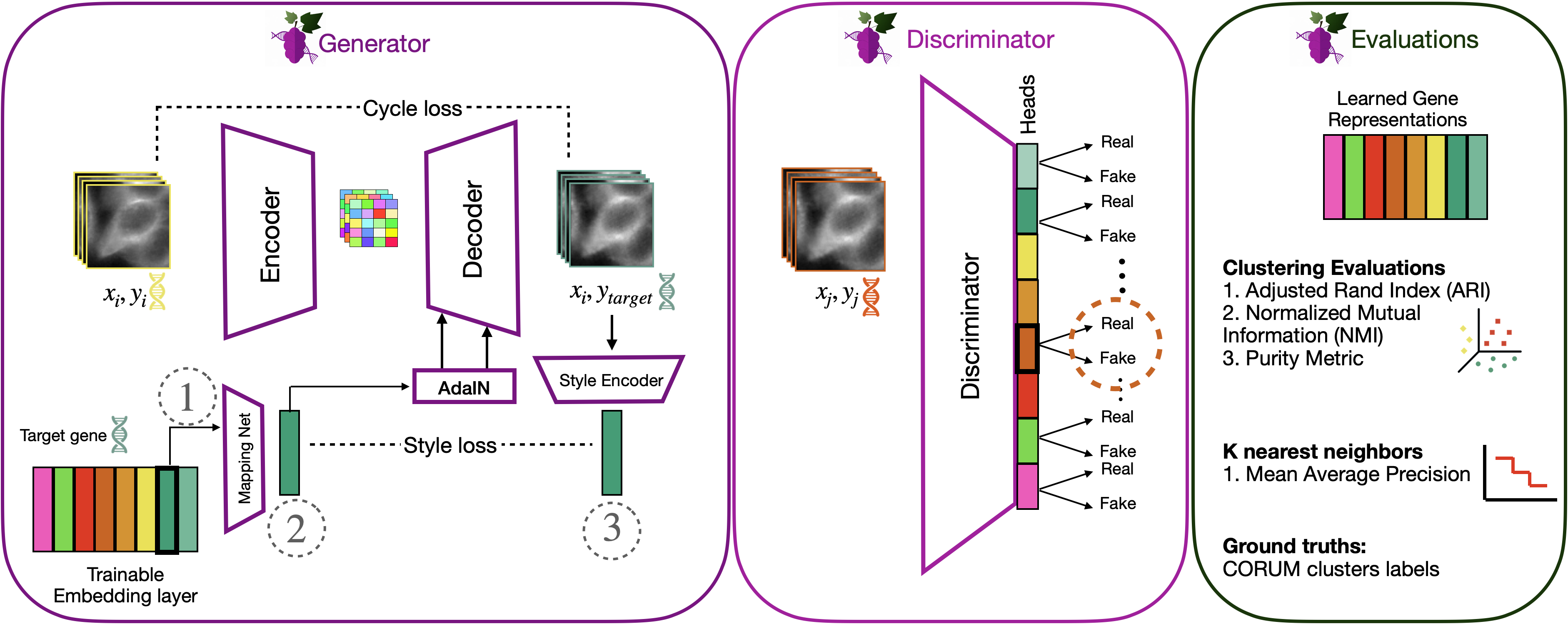}
    \caption{
\textbf{Overview of the {\includegraphics[height=1.5em]{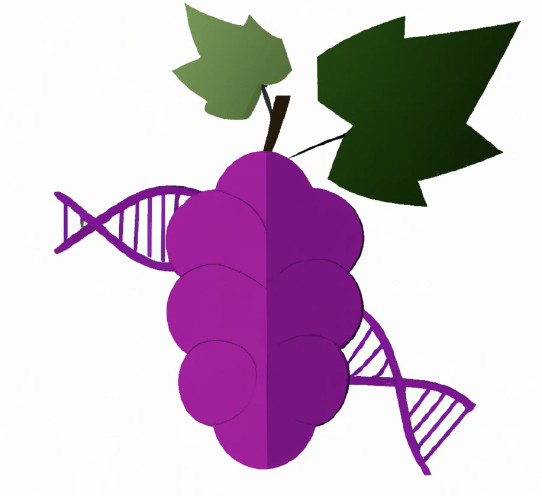}}GRAPE model}: GRAPE builds upon StarGAN~v2, and like other GAN-based models is composed of a generator and a discriminator, optimized with adversarial loss. In the GRAPE generation process, input image $x_i$ paired with its corresponding genetic perturbation $y_i$ is passed through an encoder. A target gene ($y_{target}$) is randomly selected, and its embedding is transmitted to the generator's decoder along with the encoded input, aiming to generate images reflecting the content of the input image $x_i$ and perturbation responses of  $y_{target}$. The multihead discriminator takes the generated/real image $x_j$ with its corresponding generated/real perturbation $y_j$, backpropagating the loss from real vs fake classification through the respective head of $y_j$. Our primary objective in this work is to acquire effective representations for genetic perturbations from the trainable embedding layer. Finally, we assess the quality of the representations using various evaluation metrics such as mAP and clustering metrics. For a thorough investigation, we also evaluated and comapred different potential gene representations at positions \includegraphics[height=1em]{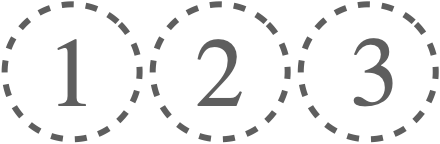}.
(Icons have been designed using images from \href{www.flaticon.com}{Flaticon.com}.)
    }
    \label{fig:pull_figure}
\end{figure*}

%
Understanding the role of genes and their functional relationship in homeostasis and disease is a fundamental challenge in biomedical research. Typical approaches to study gene function include introducing perturbations designed to destroy or interfere with specific target genes. Advances in experimentally performing gene perturbations are beginning to enable genome-scale interrogation of gene function for multiple cell types and disease states~\cite{rxrx3}. 

Automated microscopy is a powerful tool to record and quantify the often subtle cellular responses and phenotypic changes of gene perturbations, and can be combined with markers such as CellPaint~\cite{bray2016cellpaint1, gustafsdottir2013cellpaint2} for unbiased phenotypic profiling~\cite{caicedo2017data, caicedo2018weakly, caicedo2016applications}.
Conventionally, this has been done with arrayed screening, where all cells in a well receive the same perturbation; however, for large screens with millions of wells, this can become prohibitively expensive.

Recently, Optical Pooled Screening (OPS) has emerged as a cost-effective alternative to study gene function~\cite{feldman2019ops,wang2019ops, funk2022phenotypic,ramezani2023periscope,sivanandan2023posh}. Compared with arrayed screening, OPS subjects cells to a pool of perturbations that are barcoded to allow determining the specific perturbation a cell received, greatly increasing the throughput and reducing the overall cost. While this enables high perturbation throughput, a caveat is that the resulting data is at single-cell level.

A central challenge of image-based perturbation studies is analyzing the enormous datasets that are generated, in order to extract high-dimensional representations of cell state that 
capture
the biological effects of each genetic perturbation, while being invariant to numerous technical confounders.
Traditional image processing methods approach this problem by extracting a large set of predefined ``engineered" features, and then reducing dimensionality by removing noisy and highly correlated features.
Deep learning is an increasingly popular alternative to engineered features, and promises to scale with the rapidly increasing corpus of recorded data~\cite{kraus2023rxrxmae, caicedo2018weakly}; however, it may be more susceptible to overfitting technical confounders in the data
~\cite{kim2023self}.

Recently, Generative Adversarial Networks (GANs) have shown promise in addressing the impact of technical confounders on learned representations through the use of style transfer. 
In \cite{choi2020stargan}, the authors used style transfer to de-confound the training data, by generating images corresponding to the missing elements of the batch-perturbation experimental design matrix, and then train a deep learning model on the expanded training set. In addition, the authors of IMPA~\cite{palma2023predicting}, further demonstrated the use of StarGAN~v2 for generating realistic images for \textit{unseen} perturbations, for applications such as virtual screening.

Overall, we also find that style transfer~\cite{gatys2015neural} is a natural fit to the task of disentangling subtle perturbation effects from more prominent but ultimately uninformative features captured in images, such as the position of a cell within a frame. Taking inspiration from these prior works, we extend the IMPA model to extract representations of genetic perturbations (referred to interchangeably as ``style codes") from sets of cells by learning to transfer visual features between two images corresponding to different genetic perturbations, while keeping the main image content intact. The extracted style codes can be used downstream to infer relationships between perturbed genes. We refer to our method as
GRAPE: \textbf{G}ANs as \textbf{R}obust \textbf{A}dversarial \textbf{P}erturbation \textbf{E}ncoders. The model is depicted in \cref{fig:pull_figure}. 

We evaluate our learned gene embeddings based on their ability to accurately predict known functional relationships in ground truth biological datasets, and demonstrate superior performance of GRAPE embeddings compared to Gene2vec~\cite{du2019gene2vec} and IMPA~\cite{palma2023predicting} in both clustering and recall. Moreover, our approach outperforms widely used engineered features in clustering and demonstrates comparable performance in recall of biological ground truth.
Additionally, we curate and share a small biological ground truth set, to spur further innovation in this space.

%% file: sec/related_work.tex
\section{Related Work}

\subsection{Quantification of cellular phenotypes}
The phenotypic responses of cells to perturbations are often subtle and overlapping, making them difficult to distinguish even for experts. Furthermore, datasets from perturbation screens are generally prohibitively large for manual data examination. An unbiased and scalable analysis therefore requires extracting biologically informative features from images of perturbed cells. A common approach is to employ a curated bank of engineered features known to capture biologically meaningful information. Such features include size and shape of cells and cellular sub-compartments, as well as statistics and correlation of pixel intensity corresponding to relative amounts and distribution of stained biomolecules. CellProfiler~\cite{carpenter2006cellprofiler}, a reference implementation of this approach, includes a large selection of 
diverse features useful for quantifying cellular phenotypes. In the context of OPS, CellProfiler has been used to analyze a recent whole genome-wide OPS study~\cite{ramezani2023periscope}. The authors demonstrate that the obtained features cluster cells according to the biological function of perturbed genes, enabling recovery of gene interaction networks at the level of protein complexes and larger scale pathways.

Traditional image processing methods, while extracting low-level features, may fall short in capturing the entirety of biologically relevant information recorded by automated microscopes~\cite{caicedo2018weakly, rezvani2022image}. Deep learning-based computer vision is increasingly used as an alternative. A few recent OPS studies have compared the performance of engineered and learned features in different contexts. Sivanandan \textit{et al.}~\cite{sivanandan2023posh} compare engineered features mimicking CellProfiler with Vision Transformers (ViT)~\cite{dosovitskiy2020image} pretrained on ImageNet~\cite{deng2009imagenet} or trained on the OPS screen using a self-supervised objective (DINO)~\cite{caron2021emerging}. While all three approaches recover known gene interaction networks, the learned features show higher similarity of functionally related phenotypes, with the self-supervised model outperforming supervised pretraining. These results are consistent with another study comparing the performance of engineered and learned features to identify regulators of antiviral response~\cite{carlson2023genome}. The engineered features were specifically designed to measure a protein translocation bioassay reporting the cellular response to viral infection. Learned features were extracted with a pretrained convolutional neural network and an autoencoder trained from scratch. Notably, both deep learning models outperformed the engineered features, suggesting that they captured the bioassay readout better than hand-crafted features or extracted additional information not covered by prior knowledge.

%

\subsection{Style Transfer}
The goal of style transfer is to mix two images, a content image and a style reference image, to produce a new image with the content of the content image and the style of the reference image. 
In \cite{gatys2015neural}, Gatys et al. introduced a groundbreaking approach to separate and manipulate content and style in images using a pre-trained VGG model~\cite{simonyan2014very}. 
Their method aligns the output image's feature maps with the content image to preserve the content, and matches Gram matrices (covariance matrices) of feature maps between the style image and the output image to adopt the style.

In \cite{li2017demystifying}, the authors generalized the previous work by theoretically proving that matching the Gram matrices of feature maps is equivalent to minimizing the Maximum Mean Discrepancy (MMD)~\cite{gretton2012kernel} with the second order polynomial kernel, and in doing so they re-imagined style transfer as matching feature distributions between style and generated images.
Furthermore, in~\cite{li2016revisiting}, authors found that Batch Normalization (BN) layer statistics could represent style traits, and this led to Adaptive Instance Normalization (AdaIN)~\cite{huang2017arbitrary}, which is now used in popular GAN-based~\cite{goodfellow2014generative} models like the styleGAN~\cite{karras2019style} and starGAN model families~\cite{choi2018stargan, choi2020stargan}.
AdaIN takes feature statistics from a style image and applies them to a content image, effectively transferring the style of the reference image to the content image. Therefore, the input to the AdaIN layers are mean and variance embeddings and can be derived from an image or a separate encoder like a neural network. 
StarGAN~v2, a widely adopted style transfer technique employing AdaIN layers, trains a single encoder-decoder architecture capable of transferring styles across various domains, such as demographic categories.

Recent advancements have demonstrated the application of style transfer to the biological domain. Works like IST~\cite{pernice2023out} and IMPA~\cite{palma2023predicting} provide two diverse examples. IST (Interventional Style Transfer) addresses 
out-of-distribution generalization by generating counterfactual treatment response predictions in control cells.  It also paves the way for future endeavors in causal representation learning. Notably, their advancements in the style transfer pipeline involved incorporating skip connections between the encoder and decoder to uphold content preservation and implementing multiple complementary losses to discourage phenotypic alterations.

IMPA (Image Perturbation Encoder) predicts cellular morphological changes resulting from chemical and genetic perturbations given a perturbation code and an image of an unperturbed cell. The authors focus on generative performance and extrapolation to unseen perturbations. To that end, IMPA uses frozen, pre-trained perturbation embeddings (RDKit~\cite{landrum2016rdkit} and Gene2vec~\cite{du2019gene2vec} for chemical and genetic perturbations, respectively). Our work builds on IMPA which is based on StarGAN~v2, with modifications to enable learning perturbation embeddings from scratch.

%% file: sec/2_method.tex
\section{Methodology}
\label{sec:method}

\subsection{Task Definition}
The overarching goal of our model is to learn representations of optical pooled screening data that enable discovery of novel relationships between genetic perturbations.
For the purpose of methods development, we evaluate the performance of our method based on its ability to recover \textit{known} biological relationships between genes, with the underlying assumption that unsupervised recovery of known relationships is a good proxy for discovery of novel relationships.

\subsection{Data Preprocessing}
Images were first corrected for uneven illumination using a retrospective flatfield correction~\cite{lindblad2001flatfield}. Independently for each channel, pixel intensities between 0.1 and 99.9 percentiles were rescaled and clipped to range [0,1]. Single-cell images were obtained by cropping $96\times96$ pixel image patches around the cell coordinates included with the released phenotypic profiles. Pixel intensities were then normalized by channel-wise z-scoring using the mean and standard deviation of the non-targeting control images from the same experiment batch.

\begin{figure}
    \centering
    \includegraphics[width=0.9\linewidth]{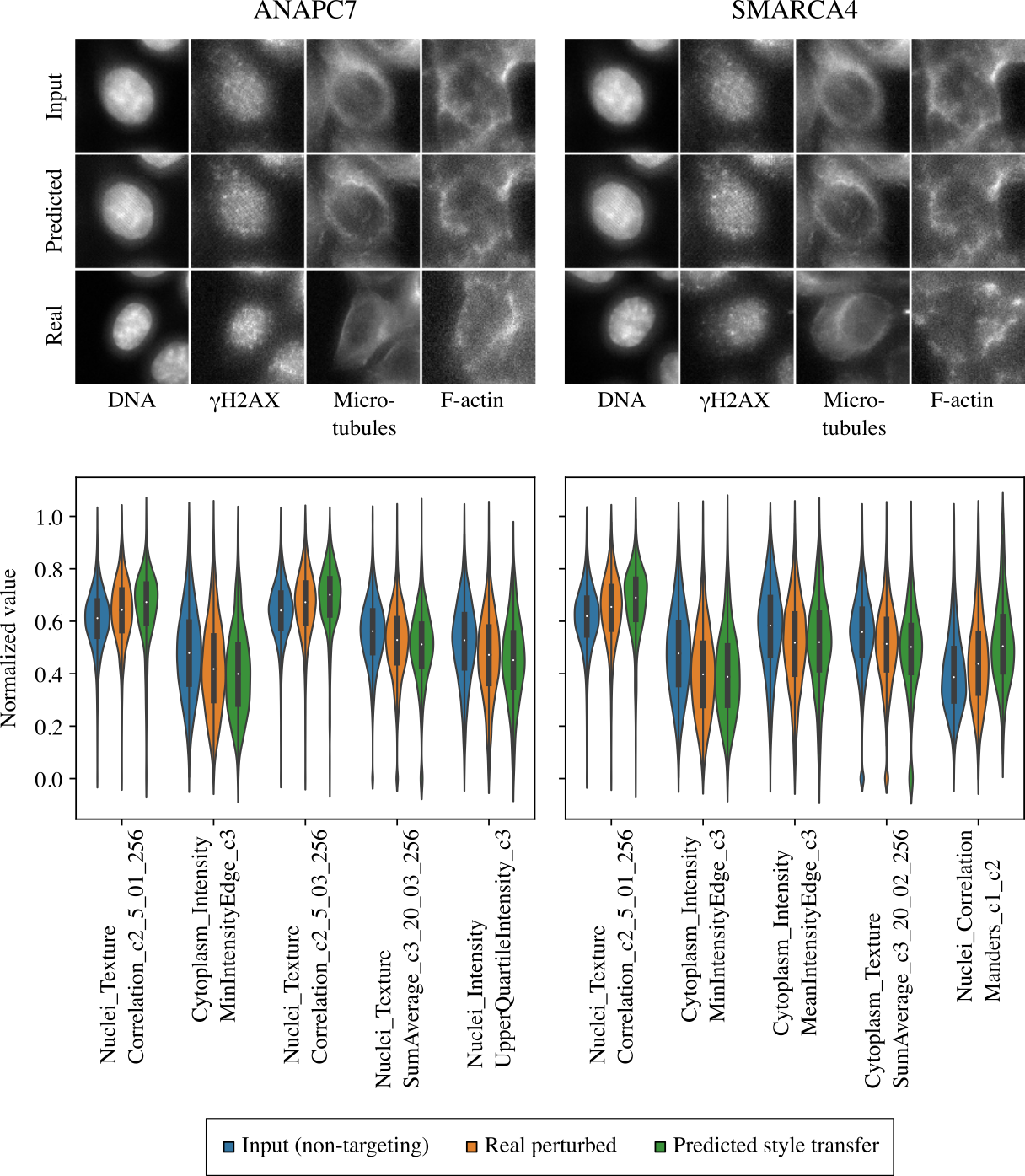}
    \caption{GRAPE generates realistic perturbation phenotypes. Top: Style transfer from an input image (non-targeting) to ANAPC7 (anaphase promoting complex subunit~7) and SMARCA4 (SWI/SNF related, matrix associated, actin dependent regulator of chromatin, subfamily~a, member~4) gene knockouts. Bottom: Distribution of five CellProfiler features most informative for classifying perturbed and control cells.}
    \label{fig:generation}
\end{figure}

\begin{figure*}
    \centering
    \includegraphics[width=0.85\textwidth]{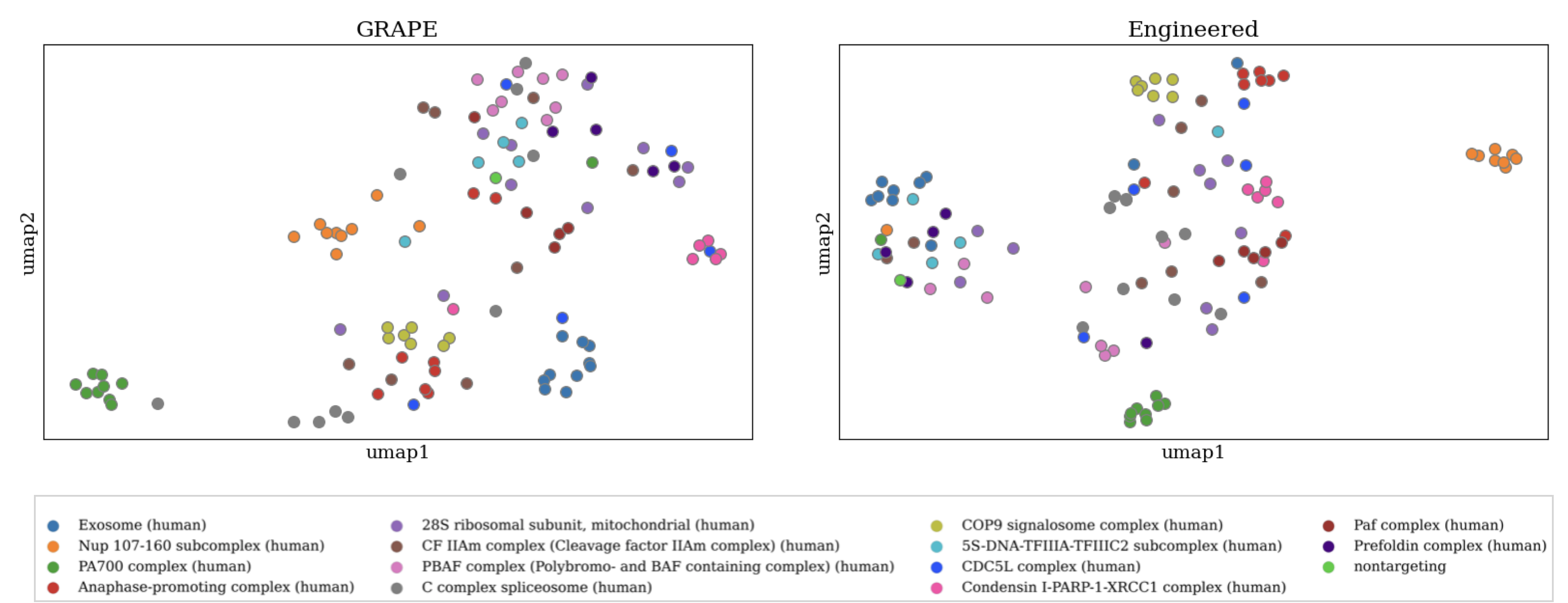}
    \caption{UMAP visualization of GRAPE representations (left), and engineered representations (right). Each dot represents a single gene, and the color indicates its ground truth CORUM cluster label.}
    \label{fig:umap}
\end{figure*}

\begin{table}
\centering
\caption{GRAPE embeddings excel across all clustering metrics when compared to the baselines and perform comparably with engineered features on the mAP metric.}
\label{tab:result}
\scalebox{0.8}{
        \begin{tabular}{@{}lp{0.02cm}cp{0.02cm}ccc@{}}  
        \toprule
         Embeddings & & & &\multicolumn{3}{c}{Clustering}\\
        \cline{5-7}
        &  &mAP && Purity & NMI & ARI\\ 
        \midrule			
        Random & & 0.104 && 0.257$\pm$0.02 & 0.310$\pm$0.02 & 0.000$\pm$0.01 \\
        Gene2vec & & 0.217 && 0.330$\pm$0.02 & 0.384$\pm$0.02 & 0.067$\pm$0.02 \\
        IMPA  & & 0.328 && 0.446$\pm$0.02& 0.478$\pm$0.02 & 0.145$\pm$0.02 \\
        Engineered & & \textbf{0.506} && \underline{0.550$\pm$0.02} & \underline{0.585$\pm$0.02} & \underline{0.253$\pm$0.03} \\
        GRAPE & & \underline{0.497} && \textbf{0.590$\pm$0.03} & \textbf{0.610$\pm$0.02} & \textbf{0.300$\pm$0.04} \\
        \bottomrule   
\end{tabular}
}
\end{table}

\subsection{Model Architecture}
Our model is a modified version of StarGAN~v2~\cite{choi2020stargan} and IMPA~\cite{palma2023predicting}, incorporating alterations to both the losses and the architecture. To train the GRAPE model, we utilize a dataset consisting of pairs ${x_i, y_i}^N_{i=1}$, where each pair comprises a single-cell image and its corresponding perturbation label. 
Our objective is to train a generator $G$ capable of producing an image that reflects the phenotypic responses associated with a given target perturbation index $z \in Z$, while retaining the content of the original image $x$, such as cell orientation. Importantly, our ultimate goal is to learn representations for each perturbation, unconfounded by nuisance features such as cell position, orientation, and technical batch effects.

Next, we discuss the modules within the GRAPE framework, walking through \cref{fig:pull_figure} from left to right.

\myparagraph{Gene Embedding Layer:}
We use a trainable embedding matrix $M$ where each of its rows represents a different genetic perturbation. This matrix begins with random normal initialization and undergoes updates throughout the training process. With our dataset containing 107 genetic perturbations, the matrix has a shape of $(107 \times d)$. We chose a length of 500 for $d$, representing the dimensionality of the genetic representations.
During the generation process, for a target perturbation $z \in \{1,2,\dots,107\}$, we extract the $z$th gene representation denoted as $M_z$ (the $z$th row of matrix $M$), and subsequently feed it into the mapping network. 
After training, we utilize this layer as our final gene embedding for downstream evaluations.

\myparagraph{Mapping Network:}
Given a latent code $M_z$, our mapping network F generates a style code $s = F(M_z)$. $F$ consists of an MLP with 3 layers in our model. We also explore different architectures for $F$, such as an attention block, and evaluate the impact of these design choices using ablation studies.

\myparagraph{Generator:}
The generator $G$ comprises an encoder and a decoder, and translates an input image $x$ into an output image $\hat{x}$, capturing the phenotypic responses associated with perturbation $z$.
The encoder computes the latent content from an image, using 3 downsampling and 2 intermediate residual blocks~\cite{he2016deep,he2016identity} with instance normalization~\cite{ulyanov2016instance}. 
The decoder consists of 2 intermediate and 3 upsampling residual blocks utilizing Adaptive Instance Normalization (AdaIN)~\cite{huang2017arbitrary}.
The perturbation style code is provided to $G$'s decoder by the mapping network $F$ through $F(M_z)$,
or by the style encoder $E$ during cycle consistency check.

\myparagraph{Discriminator:} 
The discriminator $D$ operates as a multi-task discriminator~\cite{mescheder2018training, liu2019few}, featuring multiple output branches. Each distinct branch, denoted as $D_z$, conducts binary classification, determining whether an image $x$ is a real image of perturbation $z$ or a fake image $G(x, s)$ produced by $G$.
The discriminator consists of three residual blocks, followed by two convolutional layers. The output channel of the final convolutional layer corresponds to the number of perturbations. 

\myparagraph{Style Encoder:}
Given an image $x$ and its corresponding perturbation label $y$, our encoder $E$ extracts the style code $\hat{s} = E(x)$. This block will help with the cycle loss discussed below.
The style encoder is structured with three residual blocks, followed by a convolutional layer and a fully connected layer. We adapt the architecture of StarGAN~v2 such that all parameters are shared between perturbations and there is only a single head.

\subsection{Training Objectives}
\myparagraphn{Adversarial Objective:}
During training, we randomly select a target perturbation index $z \in Z$, sample its embedding vector from matrix $M$, and generate a target style code $s = F(M_z)$. The generator $G$ takes an image $x$ (with original label $y$) and $s$ as inputs and learns to generate an output image $G(x, s)$ via an adversarial loss:
\begin{equation}
    \mathcal{L}_{adv} = \mathbb{E}_{x,y}[\log D_y(x)] + \mathbb{E}_{x,z}[1 - \log D_z(G(x,s))]
\end{equation}
Where $D_y(\cdot)$ represents the output of the $y$th branch of discriminator $D$, and similarly, $D_z(\cdot)$ denotes another branch output. The trainable matrix $M$ and the mapping network $F$ learn to provide the style code $s$ that has target perturbation $z$'s response information, and $G$ learns to apply $s$ to generate an image $G(x, s)$ that is indistinguishable from real images of the perturbation $z$.

\myparagraph{Cycle Consistency Objective:}
To ensure that the generated image $G(x,s)$ retains the perturbation-invariant traits (such as cell orientation) from its input image $x$, we implement the cycle consistency loss~\cite{kim2017learning, zhu2017unpaired, choi2018stargan}. This important loss enables us to isolate and emphasize solely the perturbation-specific information contained within the style codes and ultimately within the $M$ matrix.
\begin{equation}
     \mathcal{L}_{cyc} = \mathbb{E}_{x,y,z}[||x - G(G(x,s), \hat{s})||_1]
\end{equation}
where $\hat{s} = E(x)$ is the estimated style code of the input image $x$ with original label $y$ passed through the encoder $E$. 

\myparagraph{Style Reconstruction:}
In our adaptation of the model from StarGAN~v2, we limit the application of style loss exclusively to the cycle loss. Therefore, we restrict updates solely to the style encoder for this purpose. Our goal here is to ensure the alignment between the output of the style encoder for the generated image denoted as $E(G(x,s))$, and the style code $s$. 
\begin{equation}
    \mathcal{L}_{sty} = \mathbb{E}_{x,z}[||s\textrm{.detach()} - E(G(x,s)\textrm{.detach()})||_1]
    \label{eqn:style_loss}
\end{equation}

\myparagraph{Diversity Loss:}
In both StarGAN~v2 and IMPA~\cite{palma2023predicting}, a diversity loss was included to encourage the generator to produce diverse output images while maintaining the same content image and style code. However, in our work, we intentionally excluded both noise concatenation and the diversity loss. This decision stems from our primary focus, which centers on learning style representations, specifically related to genetic perturbations, rather than emphasizing the overall generative performance.

\myparagraph{Full objective:}
Our full objective can be written as follows:
\begin{equation}
    \underset{M,G,F,E}{\min } \underset{D}{\max} \quad \mathcal{L}_{adv} + \lambda_{cyc}\mathcal{L}_{cyc} +  \lambda_{sty}\mathcal{L}_{sty}
\end{equation}
$\lambda_{sty}$ and $\lambda_{cyc}$ are hyperparameters for each loss term.

\begin{figure*}
    \centering
    \includegraphics[width=0.85\textwidth]{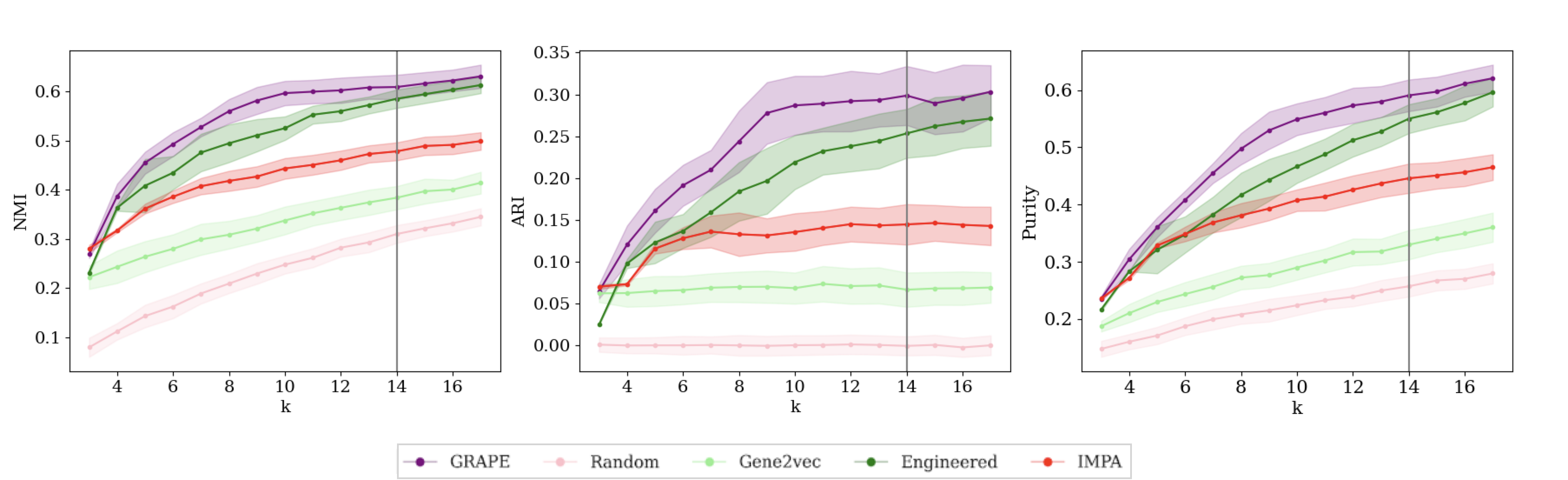}
    \caption{Performance comparison between GRAPE representations and baselines for multiple clustering metrics (left: Normalized Mutual Information, middle: Adjusted Rand Index, right: Purity) and for different numbers of clusters $k$. The ground truth number of clusters is 14, denoted by a vertical line. The $k$-means algorithm was executed 100 times for each $k$, and the standard deviation is depicted as a shaded envelope.}
    \label{fig:cluster}
\end{figure*}

\subsection{Implementation}
Input images have dimensions of $96 \times 96$ pixels and consist of 4 channels. During training, the batch size is set to 256, and we train the model for 100,000 iterations. To ensure balanced representation, the data loader employs a weighted sampler, sampling an equal number of cells per perturbation in each iteration~\cite{palma2023predicting}. On a single A100 GPU, our implementation in PyTorch~\cite{paszke2017automatic} requires approximately three days to train. During model training, we employ the non-saturating adversarial loss~\cite{goodfellow2014generative} integrated with $R1$ regularization~\cite{mescheder2018training} with $\gamma = 1$. We use an Adam optimizer~\cite{kingma2014adam} with $\beta_1 = 0$ and $\beta_2 = 0.99$. For all modules including $G$, $D$, $M$, $F$, and $E$, the learning rates and weight decays are set to 0.0001. We initialize the weights of all modules using He initialization~\cite{he2015delving}.
Both $\lambda_{sty}$ and $\lambda_{cyc}$ hyperparameters are set to 1.

\subsection{Performance Baselines}
We conducted a comparative analysis between embeddings derived from our GRAPE model's trainable gene embedding layer and Gene2vec~\cite{du2019gene2vec}. Additionally, we trained IMPA~\cite{palma2023predicting} using our dataset (see \cref{sec:dataset}) and following the original paper. This involved obtaining genetic perturbation embeddings by aggregating style embeddings from the style encoder for 500 cells per perturbation.

We also utilized engineered features from~\cite{funk2022phenotypic}, and calculated perturbation embeddings by aggregating single-cell profiles (after applying PCA) for the same set of cells used to train GRAPE.

\subsection{Evaluation Metrics}
\label{sec:eval}
We evaluate whether gene embeddings learned by GRAPE cluster according to known functional gene relationships. Using CORUM protein complexes as ground truth, we expect that perturbing genes functioning in the same protein complex should yield similar phenotypes, and thus cluster together in latent space. We selected two methods for evaluating the quality of our gene embeddings. 
The first method is mean average precision (mAP) across nearest neighbors for all genes~\cite{kim2023self}. Specifically, for a given query gene $Q$, we rank other genes by cosine distance, labeling them as $1$ if they share the same CORUM cluster as gene $Q$, and $0$ otherwise. We compute the average precision and repeat this process for all genes to derive the mAP.

%
%
The second set of metrics are based on clustering and draw inspiration from \cite{luecken2022benchmarking, kedzierska2023assessing}. These metrics include the Adjusted Rand Index (ARI)~\cite{hubert1985comparing}, Normalized Mutual Information (NMI)~\cite{pedregosa2011scikit}, and Purity metric, each focusing on different aspects of clustering. Purity evaluates the homogeneity of clusters, NMI quantifies the mutual dependence between true class labels and cluster assignments, and ARI assesses the similarity between true class labels and cluster assignments while accounting for chance.
We apply these metrics to evaluate the concordance between predicted cluster labels derived from $k$-means clustering and the ground truth cluster labels from CORUM.
Notably, we omitted the negative control embeddings from evaluation, focusing solely on perturbations with their corresponding CORUM cluster labels.
To ensure the robustness of our evaluation, we performed $k$-means clustering 100~times with random initialization for  all embeddings, including GRAPE and baseline methods. We report average and standard deviation for each metric.

\begin{table*}
\centering
\caption{Ablation results indicate that the Gene Embedding Layer (Position 1) provides the best gene representations compared to alternative layers sampled from GRAPE (\textit{i.e.,} Position 2: after the mapping network; Position 3: after the style encoder). 
Adding an attention layer does not improve model performance; however, improvements are observed when adding the cycle loss, detaching the style code in the style loss, and utilizing an exponential moving average.}
\label{tab:ablation}
\scalebox{0.8}{

        \begin{tabular}{@{}cp{0.02cm} cp{0.02cm}cp{0.02cm}cp{0.02cm}cp{0.02cm}|cp{0.02cm}ccc@{}}  
            \toprule
             & & & & & & & & & &  &  &\multicolumn{3}{c}{Clustering}\\
                          \cline{13-15}
           Position&& Attention Layer&& Detached style && Cycle Loss && EMA   &&mAP && Purity & NMI & ARI	  \\ 
        \midrule
        1&& \xmark && \cmark && \cmark && \xmark &&  0.497 &&  0.590 $\pm$ 0.03 &  0.610 $\pm$ 0.02 & \textbf{0.300 $\pm$ 0.04} 
        \\
        2&& \xmark && \cmark && \cmark && \xmark && 0.395 && 0.530 $\pm$ 0.03 &  0.561$\pm$ 0.02 &  0.231 $\pm$ 0.04 
        \\
        3&& \xmark && \cmark && \cmark && \xmark && 0.286 && 0.394 $\pm$ 0.02 & 0.445$\pm$ 0.02 &  0.010 $\pm$0.02
        \\
        1&& \cmark && \cmark && \cmark && \xmark && 0.441  && 0.540 $\pm$ 0.03 & 0.570 $\pm$ 0.03 &  0.244$\pm$ 0.04
        \\
        1&& \xmark && \xmark && \cmark && \xmark && 0.270 && 0.373 $\pm$ 0.02 & 0.422 $\pm$ 0.03 & 0.095  $\pm$ 0.02
        \\
        1&& \xmark && \cmark && \xmark && \xmark && 0.221 && 0.421 $\pm$ 0.02  & 0.446 $\pm$ 0.02 & 0.131 $\pm$0.02 
        \\
        1&& \xmark && \cmark && \cmark && \cmark && \textbf{0.498} && \textbf{0.592 $\pm$  0.03} & \textbf{0.620 $\pm$ 0.02}  & 0.286 $\pm$ 0.04 
        \\
        \bottomrule 
\end{tabular}
}
\end{table*}

%% file: sec/3_result.tex
\section{Experiments}%
\label{sec:result}%

\subsection{Dataset}
\label{sec:dataset}%
We leverage a recently published OPS study performing CRISPR knockout of 5000~essential genes in HeLa cells~\cite{funk2022phenotypic}. The released data contains 4-channel images of cells stained for biomolecular markers of DNA (DAPI), DNA-damage ($\gamma$H2AX), microtubules ($\alpha$-Tubulin) and F-actin (Phalloidin) with perturbation barcodes indicating the perturbation each cell received, which were obtained by \textit{in situ} sequencing. The authors also released a processed phenotypic profile for each cell, generated using engineered features derived from CellProfiler and other popular image processing libraries \cite{carpenter2006cellprofiler,skimg}. In their study, the authors present a detailed analysis demonstrating that these phenotypic profiles cluster according to the biological function of the perturbed genes.  We consider the released data as representative of what is achievable with engineered features (except specifically using task-specific prior knowledge if available), and thus an important baseline for our method.

For ground truth biological relationships, we use the Comprehensive Resource of Mammalian Protein Complexes (CORUM)~\cite{corum2022}. It is a database of protein complexes and the genes of their constituent proteins, covering diverse biological processes such as cell adhesion
and genetic information processing.
It is widely used to evaluate the quality of phenotypic profiles, because proteins functioning together in the same protein complex are more likely to be co-dependent, which provides a high confidence that their perturbation should yield similar phenotypes. Most other ontologies of biology include indirect and non-constitutive interactions as well as antagonists and negative regulators, which complicates their use for this specific task.

In our evaluations, we use a curated subset of CORUM that excludes protein complexes with poor intersection with the released perturbations and minimizes inter-cluster overlap~\cite{funk2022phenotypic}. From this set, we select 14~non-overlaping and functionally diverse protein complexes, each containing at least five genes. To avoid size-imbalance, we subset larger protein complexes to at most ten genes. The resulting dataset consists of 508,159~single-cell images of the following 106~perturbations, to which we add the negative control perturbations without any known target in the human genome (\textit{non-targeting}):
{\ssmall ADNP, AHCTF1, AKIRIN2, AMD1, ANKRD17, API5, ATF7IP, ATMIN, ATP6AP2, ATP6V1H, ATRX, AURKB, BRPF1, BTAF1, BTF3L4, BUD13, BYSL, C6orf15, CCNA2, CCNH, CCT4, CCT6A, CDK8, CDT1, CENPC, CENPH, CENPM, CEP55, CHERP, CHRNG, CLPS, CNOT2, COG4, COG5, COG8, COPS4, COPS5, COPS6, COQ7, CPOX, CTDNEP1, CXXC1, DDA1, DDX21, DDX39B, DIMT1, DNASE2, DRAP1, E2F3, EED, EIF1, EIF3CL, EIF6, ELL, ELP3, ENSA, EPC2, EPS15L1, ERH, ESPL1, ESPNL, EXOC1, EXOSC4, EXOSC6, EXOSC7, FAM32A, FARSA, FBL, FBXO21, FBXW11, FIBP, FKBPL, FLOT1, FNTA, FOSL1, FURIN, GFM1, GINS1, GINS3, GNB1, GPS1, GPX4, GTF2A2, GTF2B, GTF2E2, GTF3C1, ICMT, ILK, IMP3, IMPDH2, INO80, INTS1, INTS2, INTS6, INTS12, IQCF1, IRF8, JUNB, KANSL3, KAT2A, KCTD10, KPNA6, KRTAP4-5, LAMTOR1, LAMTOR3, LCN8, LDLR, LIN9, LMNB1, LPA, LSM3, LSM4, LSM6, LSM7, LTB, LUC7L3, MARK2, MCM5, MCRS1, MED18, MED23, MEPCE, MOCS3, MRPS18A, MRPS18B, MRPS33, MRPS35, MTBP, MTF1, MTOR, NACC1, NAMPT, NDUFB2, NEDD8, NFE2L1, NLE1, NMT1, NOC4L, NOL11, NOP16, NOP58, NOS1, NPEPPS, NPLOC4, NUFIP2, NUS1, PARN, PCED1B, PCF11, PDAP1, PDE4DIP, PFDN6, PHF23, PIP4K2A, PNO1, POLR1B, POLR2F, POLR2I, POLR2J3, POLR3F, POMP, PPAT, PPIA, PPP1R10, PPP4R2, PPP6R3, PREB, PRELID1, PSMA5, PSMB4, PSMD1, PSMD3, PSMD8, PSMD10, PSMD12, PSMD14, PSMG3, PWP1, QTRT1, RAD51, RBM22, RICTOR, RNASEH2A, RNF214, ROMO1, RPL7, RPL7A, RPL10A, RPL18, RPL35A, RPN1, RPS6, RPS17, RPS18, RPS24, RPS25, RPS28, RRM2, RRP15, RSL24D1, RSU1, RTCB, RXRA, SCAF11, SETD2, SETDB1, SF3B1, SF3B3, SFSWAP, SKIV2L, SLC4A7, SLC6A17, SLC7A6OS, SLC25A25, SLC25A28, SLC38A2, SLC39A10, SLTM, SLU7, SMC1A, SMC3, SMG8, SMIM7, SMNDC1, SMU1, SNRNP70, SNRPA, SNRPE, SNUPN, SP3, SRP68, SRRM2, STAG2, STX18, SUPT16H, SYMPK, TACC3, TADA3, TAF1A, TAF1B, TAF3, TAF10, TAF12, TAMM41, TCP1, TEAD3, TFAP4, TFRC, THAP11, TIMELESS, TOPBP1, TOX4, TRMT112, TSPYL5, TSSK2, TUBB, TUBB2A, TWISTNB, UBE2I, UBE2M, UCN, USP32, WDR62, WRAP73, XYLT2, YPEL5, ZC3H4, ZFC3H1, ZFP36L2, ZMAT2, ZPR1, ZRANB2}

\subsection{Evaluation}
Qualitatively, GRAPE generates realistic predictions of perturbed cells while preserving the overall context of the input cell~(\cref{fig:generation}). However, cellular phenotypes of different perturbations are often subtle and difficult to distinguish visually. To quantitatively evaluate generative performance, we style-transfered 500~non-targeting control cells to each of the perturbation styles and compared CellProfiler features of non-targeting, perturbed and predicted images. Following~\cite{palma2023predicting}, we obtained the most informative CellProfiler features by training a Random Forest classifier~\cite{ho1995random} to discriminate non-targeting from real perturbed cells and computing permutation feature importance. The distributions of the five most informative features show that style-transfer generates cells that are more similar to the target perturbation than unperturbed input cells~(\cref{fig:generation}).

As detailed in section \ref{sec:eval}, we compared learned embeddings obtained from the Gene Embedding Layer of GRAPE, with baseline methods using three clustering evaluation metrics
and a recall-based metric.
To evaluate clustering, we employ the $k$-means clustering algorithm with the parameter $k$ set to $14$, the number of ground truth clusters. 
Using the known ground truth CORUM cluster labels, we then evaluate clustering performance using purity, Adjusted Rand Index (ARI) and Normalized Mutual Information (NMI). GRAPE embeddings outperformed the baselines (random features, engineered features, IMPA and Gene2vec representations) across all three clustering metrics (see \cref{tab:result}). To ensure a comprehensive analysis, we assessed our embeddings and baseline methods across various values of $k$ in $k$-means clustering, although our dataset inherently consists of 14~clusters with 106~genes. The results depicted in \cref{fig:cluster} indicate that GRAPE embeddings outperform in all three metrics across all values of $k$.

Furthermore, we computed the mean Average Precision (mAP) for 106~genes across both GRAPE representations and other baseline methods. \cref{tab:result} shows that GRAPE embeddings outperform most baselines and are competitive with engineered features, the strongest baseline for this task.


For an additional visualization, refer to \cref{fig:umap}, which showcases UMAP embeddings of GRAPE gene representations. The colors on the plot represent the ground-truth CORUM cluster labels, demonstrating that gene perturbations within the same CORUM structure cluster together in the embedding space.

\begin{figure}
    \centering
    \includegraphics[width=.80\linewidth]{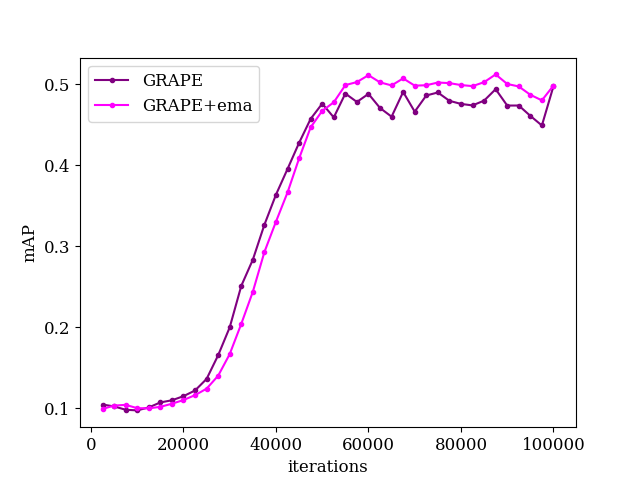}
    \caption{Comparison of mean Average Precision (mAP) for GRAPE representations, and GRAPE representation with an Exponential Moving Average (EMA) filter applied during the training process.}
    \label{fig:attn}
\end{figure}

%% file: sec/ablations.tex
\section{Ablations}
We conducted numerous ablation experiments to test which design choices are most important to GRAPE's performance.
For a comprehensive overview of all ablation results, see \cref{tab:ablation}.

\subsection{Attention}
We explored whether self-attention could capture relationships and clustering information among perturbations by introducing a single self-attention block~\cite{vaswani2017attention} before the mapping network and after the embedding layer. 
The results demonstrate that the addition of this layer did not lead to performance improvement for mAP or clustering metrics.

\subsection{Alternative Gene Embeddings}
There are multiple choices for where to extract gene embeddings from the StarGANv2 model to serve as our final gene representations for downstream tasks.
We evaluated three options: (1) the gene embedding layer, (2) the style code that is extracted by the mapping network and input to the AdaIN layer, and (3) the style code that is extracted from the generated images by the Style Encoder.
See \cref{fig:pull_figure} for the location of layers:{\includegraphics[height=1em]{figures/pos.png}} .

Interestingly, each option is qualitatively different in how the final gene representations are extracted.
For option 1, we simply extract the weights of the trained gene embedding layer, whereas for option 2, we pass the learned gene embeddings through the mapping network to generate a style code per gene.
In other words, options 1 and 2 do not require any inference-time processing.
In contrast, for option 3 we must generate cell-level style codes by passing random samples through the GRAPE model, and then aggregating the cell-level style codes returned by the Style Encoder up to the perturbation level.  To accomplish this, we randomly select 500 cells per perturbation, and average their cell-level style codes to create perturbation-level gene embeddings.
Notably, option 3 is most similar to the workflow employed by IMPA.
%
%
We compare the performance of each embedding choice after 100,000 training iterations using the mean Average Precision. Our results show that the Gene embedding layer (option 1) outperformed the other choices.

\subsection{Exponential Moving Average}
An interesting feature of our approach is that gene embeddings (and thus, genetic relationships) are extracted from the data during training, and there is no final inference step. 
However, when reviewing the training curves, we observed strong fluctuations in overall performance towards the end of training.
These fluctuations introduce an unwanted sensitivity between the model performance and the choice of stopping criteria.
Similar weight fluctuations were also observed by the authors of StarGAN~v2, and they proposed suppressing them with an exponential moving average (EMA) filter.
Thus we apply an EMA filter every 2.5k iterations with a gamma value of 0.5 and find that it both reduces sensitivity to the stopping criteria, and yields a slight improvement in mAP~(see \cref{fig:attn}).

\subsection{Losses}
We conducted experiments where we ablated different loss functions to better understand their contribution to the training process. When omitting the cycle loss during training, we observed a decrease in performance across all evaluation metrics, including mAP. In another experiment, we extended the influence of the style loss beyond the style encoder by allowing it to affect the Generator and Embedding Layer. In other words, we did not detach the style vector $s$ in the style reconstruction objective (see Equation~\ref{eqn:style_loss}). 
We also noted a drop in performance in this scenario.


%% file: sec/4_conclusion.tex
\section{Discussion}
In conclusion, our study demonstrated that style transfer through generative modeling can be used to learn high quality representations of genetic perturbations from optical pooled screening data.
We observed a significant performance improvement for all clustering metrics and competitive performance for mean Average Precision, compared to engineered features. 

It is important to acknowledge a limitation in our study, as we conducted training on a dataset comprising only 107 genes and approximately 500,000 images. We recognize the potential for enhancing the model's capabilities by extending the training to incorporate a broader range of perturbations. Future endeavors will focus on expanding this research to encompass a more diverse array of genetic perturbations, aiming to further validate and generalize the applicability of our proposed approach.

\section*{Acknowledgments}
We would like to thank Avtar Singh, Luke Funk and Rebecca Carlson for helpful discussions about OPS data.

%% file: main.bbl
\begin{thebibliography}{52}
\providecommand{\natexlab}[1]{#1}
\providecommand{\url}[1]{\texttt{#1}}
\expandafter\ifx\csname urlstyle\endcsname\relax
  \providecommand{\doi}[1]{doi: #1}\else
  \providecommand{\doi}{doi: \begingroup \urlstyle{rm}\Url}\fi

\bibitem[Bray et~al.(2016)Bray, Singh, Han, Davis, Borgeson, Hartland, Kost-Alimova, Gustafsdottir, Gibson, and Carpenter]{bray2016cellpaint1}
Mark-Anthony Bray, Shantanu Singh, Han Han, Chadwick~T Davis, Blake Borgeson, Cathy Hartland, Maria Kost-Alimova, Sigrun~M Gustafsdottir, Christopher~C Gibson, and Anne~E Carpenter.
\newblock Cell painting, a high-content image-based assay for morphological profiling using multiplexed fluorescent dyes.
\newblock \emph{Nature protocols}, 11\penalty0 (9):\penalty0 1757--1774, 2016.

\bibitem[Caicedo et~al.(2016)Caicedo, Singh, and Carpenter]{caicedo2016applications}
Juan~C Caicedo, Shantanu Singh, and Anne~E Carpenter.
\newblock Applications in image-based profiling of perturbations.
\newblock \emph{Current opinion in biotechnology}, 39:\penalty0 134--142, 2016.

\bibitem[Caicedo et~al.(2017)Caicedo, Cooper, Heigwer, Warchal, Qiu, Molnar, Vasilevich, Barry, Bansal, Kraus, et~al.]{caicedo2017data}
Juan~C Caicedo, Sam Cooper, Florian Heigwer, Scott Warchal, Peng Qiu, Csaba Molnar, Aliaksei~S Vasilevich, Joseph~D Barry, Harmanjit~Singh Bansal, Oren Kraus, et~al.
\newblock Data-analysis strategies for image-based cell profiling.
\newblock \emph{Nature methods}, 14\penalty0 (9):\penalty0 849--863, 2017.

\bibitem[Caicedo et~al.(2018)Caicedo, McQuin, Goodman, Singh, and Carpenter]{caicedo2018weakly}
Juan~C Caicedo, Claire McQuin, Allen Goodman, Shantanu Singh, and Anne~E Carpenter.
\newblock Weakly supervised learning of single-cell feature embeddings.
\newblock In \emph{Proceedings of the IEEE Conference on Computer Vision and Pattern Recognition}, pages 9309--9318, 2018.

\bibitem[Carlson et~al.(2023)Carlson, Leiken, Guna, Hacohen, and Blainey]{carlson2023genome}
Rebecca~J Carlson, Michael~D Leiken, Alina Guna, Nir Hacohen, and Paul~C Blainey.
\newblock A genome-wide optical pooled screen reveals regulators of cellular antiviral responses.
\newblock \emph{Proceedings of the National Academy of Sciences}, 120\penalty0 (16):\penalty0 e2210623120, 2023.

\bibitem[Caron et~al.(2021)Caron, Touvron, Misra, J{\'e}gou, Mairal, Bojanowski, and Joulin]{caron2021emerging}
Mathilde Caron, Hugo Touvron, Ishan Misra, Herv{\'e} J{\'e}gou, Julien Mairal, Piotr Bojanowski, and Armand Joulin.
\newblock Emerging properties in self-supervised vision transformers.
\newblock In \emph{Proceedings of the IEEE/CVF international conference on computer vision}, pages 9650--9660, 2021.

\bibitem[Carpenter et~al.(2006)Carpenter, Jones, Lamprecht, Clarke, Kang, Friman, Guertin, Chang, Lindquist, Moffat, et~al.]{carpenter2006cellprofiler}
Anne~E Carpenter, Thouis~R Jones, Michael~R Lamprecht, Colin Clarke, In~Han Kang, Ola Friman, David~A Guertin, Joo~Han Chang, Robert~A Lindquist, Jason Moffat, et~al.
\newblock Cellprofiler: image analysis software for identifying and quantifying cell phenotypes.
\newblock \emph{Genome biology}, 7:\penalty0 1--11, 2006.

\bibitem[Choi et~al.(2018)Choi, Choi, Kim, Ha, Kim, and Choo]{choi2018stargan}
Yunjey Choi, Minje Choi, Munyoung Kim, Jung-Woo Ha, Sunghun Kim, and Jaegul Choo.
\newblock Stargan: Unified generative adversarial networks for multi-domain image-to-image translation.
\newblock In \emph{Proceedings of the IEEE conference on computer vision and pattern recognition}, pages 8789--8797, 2018.

\bibitem[Choi et~al.(2020)Choi, Uh, Yoo, and Ha]{choi2020stargan}
Yunjey Choi, Youngjung Uh, Jaejun Yoo, and Jung-Woo Ha.
\newblock Stargan v2: Diverse image synthesis for multiple domains.
\newblock In \emph{Proceedings of the IEEE/CVF conference on computer vision and pattern recognition}, pages 8188--8197, 2020.

\bibitem[Deng et~al.(2009)Deng, Dong, Socher, Li, Li, and Fei-Fei]{deng2009imagenet}
Jia Deng, Wei Dong, Richard Socher, Li-Jia Li, Kai Li, and Li Fei-Fei.
\newblock Imagenet: A large-scale hierarchical image database.
\newblock In \emph{2009 IEEE conference on computer vision and pattern recognition}, pages 248--255. Ieee, 2009.

\bibitem[Dosovitskiy et~al.(2020)Dosovitskiy, Beyer, Kolesnikov, Weissenborn, Zhai, Unterthiner, Dehghani, Minderer, Heigold, Gelly, et~al.]{dosovitskiy2020image}
Alexey Dosovitskiy, Lucas Beyer, Alexander Kolesnikov, Dirk Weissenborn, Xiaohua Zhai, Thomas Unterthiner, Mostafa Dehghani, Matthias Minderer, Georg Heigold, Sylvain Gelly, et~al.
\newblock An image is worth 16x16 words: Transformers for image recognition at scale.
\newblock \emph{arXiv preprint arXiv:2010.11929}, 2020.

\bibitem[Du et~al.(2019)Du, Jia, Dai, Tao, Zhao, and Zhi]{du2019gene2vec}
Jingcheng Du, Peilin Jia, Yulin Dai, Cui Tao, Zhongming Zhao, and Degui Zhi.
\newblock Gene2vec: distributed representation of genes based on co-expression.
\newblock \emph{BMC genomics}, 20:\penalty0 7--15, 2019.

\bibitem[Fay et~al.(2023)Fay, Kraus, Victors, Arumugam, Vuggumudi, Urbanik, Hansen, Celik, Cernek, Jagannathan, Christensen, Earnshaw, Haque, and Mabey]{rxrx3}
Marta~M. Fay, Oren Kraus, Mason Victors, Lakshmanan Arumugam, Kamal Vuggumudi, John Urbanik, Kyle Hansen, Safiye Celik, Nico Cernek, Ganesh Jagannathan, Jordan Christensen, Berton~A. Earnshaw, Imran~S. Haque, and Ben Mabey.
\newblock Rxrx3: Phenomics map of biology.
\newblock \emph{bioRxiv}, 2023.

\bibitem[Feldman et~al.(2019)Feldman, Singh, Schmid-Burgk, Carlson, Mezger, Garrity, Zhang, and Blainey]{feldman2019ops}
David Feldman, Avtar Singh, Jonathan~L. Schmid-Burgk, Rebecca~J. Carlson, Anja Mezger, Anthony~J. Garrity, Feng Zhang, and Paul~C. Blainey.
\newblock Optical pooled screens in human cells.
\newblock \emph{Cell}, 179\penalty0 (3):\penalty0 787--799.e17, 2019.

\bibitem[Funk et~al.(2022)Funk, Su, Ly, Feldman, Singh, Moodie, Blainey, and Cheeseman]{funk2022phenotypic}
Luke Funk, Kuan-Chung Su, Jimmy Ly, David Feldman, Avtar Singh, Brittania Moodie, Paul~C Blainey, and Iain~M Cheeseman.
\newblock The phenotypic landscape of essential human genes.
\newblock \emph{Cell}, 185\penalty0 (24):\penalty0 4634--4653, 2022.

\bibitem[Gatys et~al.(2015)Gatys, Ecker, and Bethge]{gatys2015neural}
Leon~A Gatys, Alexander~S Ecker, and Matthias Bethge.
\newblock A neural algorithm of artistic style.
\newblock \emph{arXiv preprint arXiv:1508.06576}, 2015.

\bibitem[Goodfellow et~al.(2014)Goodfellow, Pouget-Abadie, Mirza, Xu, Warde-Farley, Ozair, Courville, and Bengio]{goodfellow2014generative}
Ian Goodfellow, Jean Pouget-Abadie, Mehdi Mirza, Bing Xu, David Warde-Farley, Sherjil Ozair, Aaron Courville, and Yoshua Bengio.
\newblock Generative adversarial nets.
\newblock \emph{Advances in neural information processing systems}, 27, 2014.

\bibitem[Gretton et~al.(2012)Gretton, Borgwardt, Rasch, Sch{\"o}lkopf, and Smola]{gretton2012kernel}
Arthur Gretton, Karsten~M Borgwardt, Malte~J Rasch, Bernhard Sch{\"o}lkopf, and Alexander Smola.
\newblock A kernel two-sample test.
\newblock \emph{The Journal of Machine Learning Research}, 13\penalty0 (1):\penalty0 723--773, 2012.

\bibitem[Gustafsdottir et~al.(2013)Gustafsdottir, Ljosa, Sokolnicki, Anthony~Wilson, Walpita, Kemp, Petri~Seiler, Carrel, Golub, Schreiber, et~al.]{gustafsdottir2013cellpaint2}
Sigrun~M Gustafsdottir, Vebjorn Ljosa, Katherine~L Sokolnicki, J Anthony~Wilson, Deepika Walpita, Melissa~M Kemp, Kathleen Petri~Seiler, Hyman~A Carrel, Todd~R Golub, Stuart~L Schreiber, et~al.
\newblock Multiplex cytological profiling assay to measure diverse cellular states.
\newblock \emph{PloS one}, 8\penalty0 (12):\penalty0 e80999, 2013.

\bibitem[He et~al.(2015)He, Zhang, Ren, and Sun]{he2015delving}
Kaiming He, Xiangyu Zhang, Shaoqing Ren, and Jian Sun.
\newblock Delving deep into rectifiers: Surpassing human-level performance on imagenet classification.
\newblock In \emph{Proceedings of the IEEE international conference on computer vision}, pages 1026--1034, 2015.

\bibitem[He et~al.(2016{\natexlab{a}})He, Zhang, Ren, and Sun]{he2016deep}
Kaiming He, Xiangyu Zhang, Shaoqing Ren, and Jian Sun.
\newblock Deep residual learning for image recognition.
\newblock In \emph{Proceedings of the IEEE conference on computer vision and pattern recognition}, pages 770--778, 2016{\natexlab{a}}.

\bibitem[He et~al.(2016{\natexlab{b}})He, Zhang, Ren, and Sun]{he2016identity}
Kaiming He, Xiangyu Zhang, Shaoqing Ren, and Jian Sun.
\newblock Identity mappings in deep residual networks.
\newblock In \emph{Computer Vision--ECCV 2016: 14th European Conference, Amsterdam, The Netherlands, October 11--14, 2016, Proceedings, Part IV 14}, pages 630--645. Springer, 2016{\natexlab{b}}.

\bibitem[Ho(1995)]{ho1995random}
Tin~Kam Ho.
\newblock Random decision forests.
\newblock In \emph{Proceedings of 3rd international conference on document analysis and recognition}, pages 278--282. IEEE, 1995.

\bibitem[Huang and Belongie(2017)]{huang2017arbitrary}
Xun Huang and Serge Belongie.
\newblock Arbitrary style transfer in real-time with adaptive instance normalization.
\newblock In \emph{Proceedings of the IEEE international conference on computer vision}, pages 1501--1510, 2017.

\bibitem[Hubert and Arabie(1985)]{hubert1985comparing}
Lawrence Hubert and Phipps Arabie.
\newblock Comparing partitions.
\newblock \emph{Journal of classification}, 2:\penalty0 193--218, 1985.

\bibitem[Karras et~al.(2019)Karras, Laine, and Aila]{karras2019style}
Tero Karras, Samuli Laine, and Timo Aila.
\newblock A style-based generator architecture for generative adversarial networks.
\newblock In \emph{Proceedings of the IEEE/CVF conference on computer vision and pattern recognition}, pages 4401--4410, 2019.

\bibitem[Kedzierska et~al.(2023)Kedzierska, Crawford, Amini, and Lu]{kedzierska2023assessing}
Kasia~Zofia Kedzierska, Lorin Crawford, Ava~Pardis Amini, and Alex~X Lu.
\newblock Assessing the limits of zero-shot foundation models in single-cell biology.
\newblock \emph{bioRxiv}, pages 2023--10, 2023.

\bibitem[Kim et~al.(2017)Kim, Cha, Kim, Lee, and Kim]{kim2017learning}
Taeksoo Kim, Moonsu Cha, Hyunsoo Kim, Jung~Kwon Lee, and Jiwon Kim.
\newblock Learning to discover cross-domain relations with generative adversarial networks.
\newblock In \emph{International conference on machine learning}, pages 1857--1865. PMLR, 2017.

\bibitem[Kim et~al.(2023)Kim, Adaloglou, Osterland, Morelli, and Zapata]{kim2023self}
Vladislav Kim, Nikolaos Adaloglou, Marc Osterland, Flavio Morelli, and Paula Andrea~Marin Zapata.
\newblock Self-supervision advances morphological profiling by unlocking powerful image representations.
\newblock \emph{bioRxiv}, pages 2023--04, 2023.

\bibitem[Kingma and Ba(2014)]{kingma2014adam}
Diederik~P Kingma and Jimmy Ba.
\newblock Adam: A method for stochastic optimization.
\newblock \emph{arXiv preprint arXiv:1412.6980}, 2014.

\bibitem[Kraus et~al.(2023)Kraus, Kenyon-Dean, Saberian, Fallah, McLean, Leung, Sharma, Khan, Balakrishnan, Celik, Sypetkowski, Cheng, Morse, Makes, Mabey, and Earnshaw]{kraus2023rxrxmae}
Oren Kraus, Kian Kenyon-Dean, Saber Saberian, Maryam Fallah, Peter McLean, Jess Leung, Vasudev Sharma, Ayla Khan, Jia Balakrishnan, Safiye Celik, Maciej Sypetkowski, Chi~Vicky Cheng, Kristen Morse, Maureen Makes, Ben Mabey, and Berton Earnshaw.
\newblock Masked autoencoders are scalable learners of cellular morphology.
\newblock 2023.

\bibitem[Landrum et~al.(2016)]{landrum2016rdkit}
Greg Landrum et~al.
\newblock Rdkit: Open-source cheminformatics software.
\newblock 2016.

\bibitem[Li et~al.(2016)Li, Wang, Shi, Liu, and Hou]{li2016revisiting}
Yanghao Li, Naiyan Wang, Jianping Shi, Jiaying Liu, and Xiaodi Hou.
\newblock Revisiting batch normalization for practical domain adaptation.
\newblock \emph{arXiv preprint arXiv:1603.04779}, 2016.

\bibitem[Li et~al.(2017)Li, Wang, Liu, and Hou]{li2017demystifying}
Yanghao Li, Naiyan Wang, Jiaying Liu, and Xiaodi Hou.
\newblock Demystifying neural style transfer.
\newblock \emph{arXiv preprint arXiv:1701.01036}, 2017.

\bibitem[Lindblad and Bengtsson(2001)]{lindblad2001flatfield}
Joakim Lindblad and Ewert Bengtsson.
\newblock A comparison of methods for estimation of intensity non uniformities in 2d and 3d microscope images of fluorescence stained cells.
\newblock In \emph{Proceedings of the Scandinavian Conference On Image Analysis}, pages 264--271, 2001.

\bibitem[Liu et~al.(2019)Liu, Huang, Mallya, Karras, Aila, Lehtinen, and Kautz]{liu2019few}
Ming-Yu Liu, Xun Huang, Arun Mallya, Tero Karras, Timo Aila, Jaakko Lehtinen, and Jan Kautz.
\newblock Few-shot unsupervised image-to-image translation.
\newblock In \emph{Proceedings of the IEEE/CVF international conference on computer vision}, pages 10551--10560, 2019.

\bibitem[Luecken et~al.(2022)Luecken, B{\"u}ttner, Chaichoompu, Danese, Interlandi, M{\"u}ller, Strobl, Zappia, Dugas, Colom{\'e}-Tatch{\'e}, et~al.]{luecken2022benchmarking}
Malte~D Luecken, Maren B{\"u}ttner, Kridsadakorn Chaichoompu, Anna Danese, Marta Interlandi, Michaela~F M{\"u}ller, Daniel~C Strobl, Luke Zappia, Martin Dugas, Maria Colom{\'e}-Tatch{\'e}, et~al.
\newblock Benchmarking atlas-level data integration in single-cell genomics.
\newblock \emph{Nature methods}, 19\penalty0 (1):\penalty0 41--50, 2022.

\bibitem[Mescheder et~al.(2018)Mescheder, Geiger, and Nowozin]{mescheder2018training}
Lars Mescheder, Andreas Geiger, and Sebastian Nowozin.
\newblock Which training methods for gans do actually converge?
\newblock In \emph{International conference on machine learning}, pages 3481--3490. PMLR, 2018.

\bibitem[Palma et~al.(2023)Palma, Theis, and Lotfollahi]{palma2023predicting}
Alessandro Palma, Fabian~J Theis, and Mohammad Lotfollahi.
\newblock Predicting cell morphological responses to perturbations using generative modeling.
\newblock \emph{bioRxiv}, pages 2023--07, 2023.

\bibitem[Paszke et~al.(2017)Paszke, Gross, Chintala, Chanan, Yang, DeVito, Lin, Desmaison, Antiga, and Lerer]{paszke2017automatic}
Adam Paszke, Sam Gross, Soumith Chintala, Gregory Chanan, Edward Yang, Zachary DeVito, Zeming Lin, Alban Desmaison, Luca Antiga, and Adam Lerer.
\newblock Automatic differentiation in pytorch.
\newblock 2017.

\bibitem[Pedregosa et~al.(2011)Pedregosa, Varoquaux, Gramfort, Michel, Thirion, Grisel, Blondel, Prettenhofer, Weiss, Dubourg, et~al.]{pedregosa2011scikit}
Fabian Pedregosa, Ga{\"e}l Varoquaux, Alexandre Gramfort, Vincent Michel, Bertrand Thirion, Olivier Grisel, Mathieu Blondel, Peter Prettenhofer, Ron Weiss, Vincent Dubourg, et~al.
\newblock Scikit-learn: Machine learning in python.
\newblock \emph{the Journal of machine Learning research}, 12:\penalty0 2825--2830, 2011.

\bibitem[Pernice et~al.(2023)Pernice, Doron, Quach, Pratapa, Kenjeyev, De~Veaux, Hirano, and Caicedo]{pernice2023out}
Wolfgang~M Pernice, Michael Doron, Alex Quach, Aditya Pratapa, Sultan Kenjeyev, Nicholas De~Veaux, Michio Hirano, and Juan~C Caicedo.
\newblock Out of distribution generalization via interventional style transfer in single-cell microscopy.
\newblock In \emph{Proceedings of the IEEE/CVF Conference on Computer Vision and Pattern Recognition}, pages 4325--4334, 2023.

\bibitem[Ramezani et~al.(2023)Ramezani, Bauman, Singh, Weisbart, Yong, Lozada, Way, Kavari, Diaz, Haghighi, Batista, P{\'e}rez-Schindler, Claussnitzer, Singh, Cimini, Blainey, Carpenter, Jan, and Neal]{ramezani2023periscope}
Meraj Ramezani, Julia Bauman, Avtar Singh, Erin Weisbart, John Yong, Maria Lozada, Gregory~P. Way, Sanam~L. Kavari, Celeste Diaz, Marzieh Haghighi, Thiago~M. Batista, Joaqu{\'\i}n P{\'e}rez-Schindler, Melina Claussnitzer, Shantanu Singh, Beth~A. Cimini, Paul~C. Blainey, Anne~E. Carpenter, Calvin~H. Jan, and James~T. Neal.
\newblock A genome-wide atlas of human cell morphology.
\newblock \emph{bioRxiv}, 2023.

\bibitem[Rezvani et~al.(2022)Rezvani, Bigverdi, and Rohban]{rezvani2022image}
Arghavan Rezvani, Mahtab Bigverdi, and Mohammad~Hossein Rohban.
\newblock Image-based cell profiling enhancement via data cleaning methods.
\newblock \emph{Plos one}, 17\penalty0 (5):\penalty0 e0267280, 2022.

\bibitem[Simonyan and Zisserman(2014)]{simonyan2014very}
Karen Simonyan and Andrew Zisserman.
\newblock Very deep convolutional networks for large-scale image recognition.
\newblock \emph{arXiv preprint arXiv:1409.1556}, 2014.

\bibitem[Sivanandan et~al.(2023)Sivanandan, Leitmann, Lubeck, Sultan, Stanitsas, Ranu, Ewer, Mancuso, Phillips, Kim, et~al.]{sivanandan2023posh}
Srinivasan Sivanandan, Bobby Leitmann, Eric Lubeck, Mohammad~Muneeb Sultan, Panagiotis Stanitsas, Navpreet Ranu, Alexis Ewer, Jordan~E Mancuso, Zachary~F Phillips, Albert Kim, et~al.
\newblock A pooled cell painting crispr screening platform enables de novo inference of gene function by self-supervised deep learning.
\newblock \emph{bioRxiv}, pages 2023--08, 2023.

\bibitem[Tsitsiridis et~al.(2022)Tsitsiridis, Steinkamp, Giurgiu, Brauner, Fobo, Frishman, Montrone, and Ruepp]{corum2022}
George Tsitsiridis, Ralph Steinkamp, Madalina Giurgiu, Barbara Brauner, Gisela Fobo, Goar Frishman, Corinna Montrone, and Andreas Ruepp.
\newblock {CORUM: the comprehensive resource of mammalian protein complexes–2022}.
\newblock \emph{Nucleic Acids Research}, 51\penalty0 (D1):\penalty0 D539--D545, 2022.

\bibitem[Ulyanov et~al.(2016)Ulyanov, Vedaldi, and Lempitsky]{ulyanov2016instance}
Dmitry Ulyanov, Andrea Vedaldi, and Victor Lempitsky.
\newblock Instance normalization: The missing ingredient for fast stylization.
\newblock \emph{arXiv preprint arXiv:1607.08022}, 2016.

\bibitem[van~der Walt et~al.(2014)van~der Walt, Schönberger, Nunez-Iglesias, Boulogne, Warner, Yager, Gouillart, Yu, and {scikit-image contributors}]{skimg}
S van~der Walt, JL Schönberger, J Nunez-Iglesias, F Boulogne, JD Warner, N Yager, E Gouillart, T Yu, and {scikit-image contributors}.
\newblock scikit-image: image processing in python.
\newblock \emph{PeerJ}, 2:\penalty0 e453, 2014.

\bibitem[Vaswani et~al.(2017)Vaswani, Shazeer, Parmar, Uszkoreit, Jones, Gomez, Kaiser, and Polosukhin]{vaswani2017attention}
Ashish Vaswani, Noam Shazeer, Niki Parmar, Jakob Uszkoreit, Llion Jones, Aidan~N Gomez, {\L}ukasz Kaiser, and Illia Polosukhin.
\newblock Attention is all you need.
\newblock \emph{Advances in neural information processing systems}, 30, 2017.

\bibitem[Wang et~al.(2019)Wang, Lu, Emanuel, Babcock, and Zhuang]{wang2019ops}
C Wang, T Lu, G Emanuel, HP Babcock, and X Zhuang.
\newblock Imaging-based pooled crispr screening reveals regulators of lncrna localization.
\newblock \emph{Proc Natl Acad Sci U S A}, 116:\penalty0 10842--10851, 2019.

\bibitem[Zhu et~al.(2017)Zhu, Park, Isola, and Efros]{zhu2017unpaired}
Jun-Yan Zhu, Taesung Park, Phillip Isola, and Alexei~A Efros.
\newblock Unpaired image-to-image translation using cycle-consistent adversarial networks.
\newblock In \emph{Proceedings of the IEEE international conference on computer vision}, pages 2223--2232, 2017.

\end{thebibliography}
